# CONSTRUCTING THE GÖDEL UNIVERSE

Rainer Burghardt[1]

By a suitable transformation, we derive the rotating Gödel universe from a static one and we show, how rotation may be implemented geometrically. The rotation law turns out to be a differential one. By increasing distance from the rotation axis the velocity of a rotating point will exceed the velocity of light and the cosmos has a cut-off radius. Thus, closed time-like curves do not occur in the Gödel universe.

## 1. INTRODUCTION

As many rotating systems in gravitation theory were found, the question arises how the effects of rotation are geometrically implemented in the theory. Rotation is recognized in the metric by an oblique angled coordinate system, in general with respect to two coordinates. As one assumes that such a metric describes an embedded $V_4$ in a higher dimensional flat space, it is obvious to search for a surface belonging to that metric. The Gödel metric [1] has been of considerable interest for many authors, but the problem of embedding does not seem to have been solved up to date. J. ROSEN [2] has presented a 10-dimensional embedding, but this is a local one and does not give any information about a surface.

Therefore, we are searching for other possibilities. We start with a static metric and define a helix twisting around the time-like coordinate axis. This will be the world line of an observer, comoving with the rotating matter of the Gödel universe. If the static metric is subjected to a suitable change of the coordinates, it will transform to the Gödel metric and one of the new coordinate lines will be the helix mentioned above.

The field strength of rotation is constant and of order $1/\mathcal{R}$, where $\mathcal{R}$ is the radius of a pseudo-sphere, but the angular velocity is a function of the radial coordinate. The velocity of the rotating matter is constant on cylinders with the z-coordinate as symmetry axis, but has different values in neighboring cylinders. That implies a *differential rotation law*. As the velocity on the orbits is increasing with the distance from the rotation axis, it is exceeding the velocity of light at the *cut-off radius*. Beyond this radius, the Gödel metric has 'closed time-like curves'. Since physics breaks down in that region, no reality can be assigned to that lines and there is no chance for time travels.

## 2. FROM STATIC TO STATIONARY METRIC

The static metric

$$ds^2 = \mathcal{R}^2 d\psi^2 + dz^2 + \mathcal{R}^2 sh^2\psi\, d\varphi^2 - dt^2 \tag{2.1}$$

describes a pseudo-sphere $x^\alpha x^\alpha = \mathcal{R}^2$, $\alpha = 0,1,2$, parametrized by

$$\begin{aligned}x^0 &= \mathcal{R}\, ch\psi \\ x^1 &= i\mathcal{R}\, sh\psi\, cos\varphi \\ x^2 &= i\mathcal{R}\, sh\psi\, sin\varphi\end{aligned} \tag{2.2}$$

and is cylindrical in z and pseudo-cylindrical in t. Using cylindrical coordinates ($\psi = 1, z = 2, \varphi = 3, it = 4$), we read off from (2.1) the tetrads

$$\overset{1}{e}_1 = \mathcal{R},\quad \overset{2}{e}_2 = 1,\quad \overset{3}{e}_3 = \sigma = \mathcal{R}\, sh\psi,\quad \overset{4}{e}_4 = 1, \tag{2.3}$$

and we easily get the only non-vanishing components of the connexion coefficients $A_{mn}{}^s$ in tetrad representation

$$C_1 = A_{31}{}^3 = -A_{33}{}^1 = \frac{1}{\sigma}\sigma_{/1} = \frac{1}{\mathcal{R}} cth\psi, \tag{2.4}$$

which explain the curvature of the surface in the [1,3]-section. The Ricci-tensor has the simple form

$$\begin{aligned}R_{mn} &= -\left[C_{n/m} + C_m C_n\right] - c_m c_n \left[C^s{}_{/s} + C^s C_s\right] \\ R_{11} &= R_{33} = -\frac{1}{\mathcal{R}^2},\quad R_{22} = R_{44} = 0,\quad R = -\frac{2}{\mathcal{R}^2}. \\ c_m &= \overset{3}{e}_m,\quad m = 1,...,4\end{aligned} \tag{2.5}$$

The transformation

$$e_{m'}^i = A_{m'}^m e_m^i,\quad A_{3'}^3 = A_3^{3'} = 1,\quad A_{3'}^4 = -A_3^{4'} = ith\Theta,\quad A_{4'}^4 = A_4^{4'} = 1 \tag{2.6}$$

rotates the third vector of the unit tetrads $e_m^i$ on the surface of the pseudo-cylinder through an angle $\Theta$. In general $\Theta$ is a function of the coordinates, but constant on a specific cylindrical surface. The length of the new vectors $e_{3'}^i$ is $\tau = 1/ch\Theta$ and they are tangent to a helix parametrized by

$$\begin{aligned}x &= \sigma\, cos\varphi \\ y &= \sigma\, sin\varphi, \\ w &= b\varphi\end{aligned} \tag{2.7}$$

where $b = \sigma ith\Theta$ is the pitch of the helix and w is pointing into the direction of the coordinate time. The line element $ds^2 = (\sigma^2 + b^2)d\varphi^2$ may also be written as $ds = \sigma d\varphi / ch\Theta$. The contravariant components of a vector are derived by dividing the projection of the quantity on the axis by the corresponding value of the tangent tetrad vector: $dx^{3'} = ds/\underset{3'}{\tau}$. Thus, we have in agreement with (2.6)

$$dx^{3'} = dx^3, \ dx^{4'} = dx^4 + dw = dx^4 + ith\Theta dx^3 \tag{2.8}$$

With the identification

$$\omega\sigma = th\Theta, \omega = \omega(\psi), \sigma = \sigma(\psi), \tag{2.9}$$

where ω is the angular velocity and ωσ the velocity of the rotating matter, (2.8) reads as

$$dx^{3'} = \sigma d\varphi, \ dt' = \omega\sigma \cdot \sigma d\varphi + dt. \tag{2.10}$$

By a coordinate transformation $\overset{m}{e}_{i'} = A_{i'}^{i} \overset{m}{e}_{i}$, related to (2.6), we get the tetrad (2.3) in the oblique angled coordinate system

$$\underset{3'}{e^3} = \sigma, \underset{3'}{e^4} = i\omega\sigma \cdot \sigma, \underset{4'}{e^4} = 1, \underset{3}{e^{3'}} = \frac{1}{\sigma}, \underset{3}{e^{4'}} = -i\omega\sigma, \underset{4}{e^{4'}} = 1, \tag{2.11}$$

and the metric (2.1) appears in a stationary form:

$$ds^2 = \mathcal{R}^2 d\psi^2 + dz^2 + \sigma^2 d\varphi^2 - [\omega\sigma \cdot \sigma d\varphi + dt]^2. \tag{2.12}$$

## 3. PHYSICAL INTERPRETATION OF THE STATIONARY METRIC

If the transition from a static metric to a stationary one by a transformation of the coordinates is not a mere change of the representation, but should have effects that admit a physical interpretation, one has to assign to the new vector field $\underset{3'}{e^i}$ invariant properties.

This field, a function of the radial coordinate, defines a new structure on the $V_4$ with the metric (2.1), exceeding the properties of the Riemannian space. The Ricci tensor in tetrad components includes the object of anholonomity due to the curl of this field. TREDER has treated the problem of geometrized electromagnetic potentials in the framework of an 'already unified' theory [3,4]. His methods are in some relation to our approach.

Writing down the metric (2.12) in the form

$$ds^2 = \mathcal{R}^2 d\psi^2 + dz^2 + (1 - \omega^2\sigma^2) \ \sigma^2 d\varphi^2 - 2\omega\sigma^2 d\varphi\, dt - dt^2 \tag{3.1}$$

we find that for $t=0$ and $\omega\sigma > 1$ (this would be equivalent to $th\Theta > 1$!) the curves related to $\varphi$ become time-like. This is the case, when the velocity of the rotating matter is exceeding the velocity of light (c=1) at the distance

$$\sigma_c = 1/\omega_c \tag{3.2}$$

from the rotation axis. The region beyond the cut-off radius $\sigma_c$ is unphysical, and any model of the type (3.1) is unphysical too, if we do not choose for $\omega = \omega(\psi)$ a function, so that $\omega\sigma \to k$, $(k<1)$ for $\sigma \to \infty$. There are no physically relevant closed time-like lines nor is there any possibility for time journeys. The Gödel model is a special case of (3.1), if we set

$$\omega\sigma = \sqrt{2}\,th\chi. \tag{3.3}$$

Renaming the variables $\mathbb{R} \to a$, $\psi \to 2r$, $z \to 2ay$, $t \to 2at$; $\psi = 2\chi$ in (3.1), we get the Gödel metric

$$ds^2 = 4a^2\left[dr^2 + dy^2 - (sh^4 r - sh^2 r)d\varphi^2 - 2\sqrt{2}\,sh^2 r\, d\varphi\, dt - dt^2\right], \tag{3.4}$$

and Gödel's condition for time-like lines turns out to be the same as our condition (3.2). Instead of (3.3) there are other possibilities to implement a law of rotation, but Gödel's theory leads to a constant field strength of rotation, proportional to $1/\mathbb{R}^2$, which simplifies the field equations considerably, as we will later see. A suitable ansatz like $\omega \sim 1/\sigma^n$, $n \geq 2$ avoids the problem of cut-off radius but demands a construction for the energy tensor, which can hardly be explained.

## 4. THE FIELD EQUATIONS

Under the coordinate transformation ($i \to i'$) changing (2.1) to the Gödel metric (3.1), the tetrad components of the connexion

$$A_{nm}{}^s = \overset{s}{e_i}\, e^i{}_{[m/n]} + g^{sr} g_{nt}\, \overset{t}{e_i}\, e^i{}_{[m/r]} + g^{sr} g_{mt}\, \overset{t}{e_i}\, e^i{}_{[n/r]}, \quad g_{mn} = \delta_{mn} \tag{4.1}$$

transform inhomogeneously and get a new contribution

$$H_{nm} u^s + H^s{}_n u_m + H^s{}_m u_n, \tag{4.2}$$

where $u_m = \{0,0,0,1\}$ are the components of the 4-velocity of the rotating matter in an orthogonal local system and the $H_{nm}$ are the components of the force of rotation. They may be calculated directly from (4.1) and (2.11) by

$$H_{13} = \frac{1}{2}\left[\overset{4}{e_{3'}} e^{3'}{}_{3/1} + \overset{4}{e_{4'}} e^{4'}{}_{3/1}\right] = -i\left[\omega\sigma_{/1} + \frac{1}{2}\omega_{/1}\sigma\right]. \tag{4.3}$$

The first term is $\omega\, ch\psi$ and corresponds to the classical Coriolis force, but the second has its origin in the differential rotation law

$$\omega = \omega(\chi) = \frac{1}{\sqrt{2}\mathcal{R}\,ch^2\chi} \quad , \quad \frac{1}{2}\omega_{/1}\sigma = -\omega\,sh^2\chi \,. \tag{4.4}$$

The velocity of the rotating matter relative to the static system (2.3) is a function of the radial coordinate: $\omega(\chi)\sigma(\chi) = \sqrt{2}\,th\chi$ and is increasing with the distance $\sigma$ from the rotation axis $z$, but constant on the cylinders $\sigma = const$. Neighboring cylinders are rotating with different velocities. Observers on locally non-rotating cylinders will experience deformation forces, which are easy to calculate from (4.1), inserting locally non-rotating tetrads [5]. The new tetrads are derived from (2.11) by the generalized Lorentz transformation

$$A_3^3 = A_4^4 = ch\Theta = \alpha,\ A_3^4 = -A_4^3 = ish\Theta = i\alpha\omega\sigma,\ \alpha = 1/\sqrt{1-\omega^2\sigma^2}\,, \tag{4.5}$$

but rearranging the metric (3.1) to

$$ds^2 = \mathcal{R}^2 d\psi^2 + dz^2 + \left(\alpha^{-1}\sigma\,d\varphi - \alpha\omega\sigma\,dt\right)^2 - \alpha^2 dt^2 \tag{4.6}$$

yields the same result.

The only non-vanishing component of $H_{nm}$ is constant

$$H_{13} = -\frac{i}{\sqrt{2}\mathcal{R}} \tag{4.7}$$

and the square $H^2 = H_{nm}H^{nm} = -1/\mathcal{R}^2$ is essential for the field equations. The Ricci tensor includes rotational effects

$$R_{mn} = -\left[C_{n/m} + C_m C_n\right] - c_m c_n \left[C^s{}_{/s} + C^2\right] + 2u_{(m}H^s{}_{n)/s} + 2H_m{}^r H_{nr} + u_m u_n H^2 \tag{4.8}$$

and the field equations

$$R_{mn} - \frac{1}{2}R g_{mn} - \lambda g_{mn} = -\kappa T_{mn} \tag{4.9}$$

are satisfied with $\lambda = -R/2,\ T_{mn} = \mu_0 u_m u_n,\ \kappa\mu_0 = 1/\mathcal{R}^2$. Since

$$H^s{}_{n/s} = 0\,, \tag{4.10}$$

there is no coupling to quadratic terms. The brackets above have the value $1/\mathcal{R}^2$ as shown by (2.5). The field equations get rather simple and no impressing field mechanism can be read off from (4.8, 4.9).

## 5. ANOTHER IMPLEMENTATION OF ROTATION

Another possibility to start from (2.1) or any other static metric is to retain the orthogonal coordinate system (i), but to subject the tetrad (2.3) to the generalized Lorentz transformation (4.5). Such a transformation produces a set of rotating frames in a static universe, but this rotation is a relative one. One is not able to decide, whether the universe is at rest and the observers are rotating or whether the universe is rotating relatively to these observers. In the first case centrifugal and Coriolis forces acting on Newton's bucket are produced by the rotation of the system, in agreement with the classical theory of rotation. In the second case the rotating universe transports rotational field energy. This energy has its origin in the quadratic terms of the field strengths. This field energy generates a negative mass in the center of rotation, which has a repulsive action. These mechanisms have been described by HUND [6] in detail and discussed by BURGHARDT [7] with respect to the Sagnac experiment. Rotational field forces shorten and lengthen the optical paths of the light beams on the Sagnac platform and cause a shift of the spectral lines in the interferometer, while the velocity of light remains constant, although the system is accelerated. Thus, Sagnac's experiment is the pendant to the Michelson experiment, also concerning its interpretation.

How the contribution of the rotating reference system to the field equations are to be treated, was pointed out by BURGHARDT [7] for a flat model. The new model is dynamically endowed with the relativistic Coriolis force and the centrifugal force

$$H_{\alpha 3} = -i\alpha^2 \omega \sigma_{/\alpha}, \; F_\alpha = \alpha^2 \omega^2 \sigma \sigma_{/\alpha}, \; \alpha = 1,2,3 \;, \tag{5,1}$$

which decouple from the field equations and satisfy Maxwell-like equations

$$\begin{aligned} rot\, \boldsymbol{H} + 2\, \boldsymbol{H} \times \boldsymbol{F} &= 0, \quad div\, \boldsymbol{H} = 0 \\ div\, \boldsymbol{F} - (\, \boldsymbol{H}^2 - \boldsymbol{F}^2\, ) &= 0, \quad rot\, \boldsymbol{F} = 0 \end{aligned} \;, \tag{5.2}$$

where *rot* and *div* are 3-dimensinal covariant differential operators. Moreover, the Pointing vector and the field energy are conserved

$$div(\, 2\boldsymbol{H} \times \boldsymbol{F}\, ) = 0, \; \frac{\partial}{\partial t}(\, \boldsymbol{H}^2 - \boldsymbol{F}^2\, ) = 0 \;. \tag{5.3}$$

To our knowledge, quite a lot of rotating models exhibit such a structure, if reformulated in tetrad formalism and if their field strengths are separated carefully. We suggest a systematic search for known models that have a hidden dynamical implementation of rotation by a generalized Lorentz transformation[1]. In this case, rotation is not a geometrical property of Riemannian space, but due to a local tetrad transformation in the tangential space.

# CONCLUSIONS

Usually one tries to understand the physical background of the metric in calculating the geodesics. In the present paper we obviously get more information by consequently using tetrad formalism and by properly arranging the components of the connexion for geometrical and physical interpretation. The Gödel metric exhibits some features which let us assume that apart from the mathematical importance the possibilities for physical interpretation are not really amazing. There is no centrifugal and no inertial force. The velocity of the matter is increasing with the radial distance and exceeding the velocity of light at the cut-off radius, beyond which physical interpretation is impossible. In this region 'closed time-like geodesics' occur. The angular velocity is a function of the radial coordinate, but does not decrease fast enough to avoid the existence of a cut-off radius.

---

[1] The model of LEWIS [8,9] can even be transformed to the static Weil metric by a holonomic coordinate transformation.

## ACKNOWLEDGEMENTS

I am indebted to Prof. H.-J. Treder for his kind interest in this work.

---

[1] A-2061, Obritz 246; Homepage http://arg.at.tf, email: arg@i-one.at